\documentstyle[pra,aps,psfig]{revtex}
\def \<{\langle}
\def \>{\rangle}
\begin{document}
\draft
\twocolumn[\hsize\textwidth\columnwidth\hsize\csname @twocolumnfalse\endcsname
\title{A numerical approach to the ground and excited states of a Bose-Einstein
condensed gas confined in a completely anisotropic trap}
\author{B. I. Schneider}
\address{Physics Division, National Science Foundation, Arlington,
Virginia 22230}
\author{D. L. Feder}
\address{University of Oxford, Parks Road, Oxford OX1 3PU, U.K. and 
          National Institute of Standards and Technology, Gaithersburg, MD 
          20899}
\date{\today}
\maketitle
\begin{abstract}
The ground and excited states of a weakly interacting and dilute Bose-Einstein
condensed gas, confined in a completely anisotropic harmonic oscillator
potential, are determined at zero temperature within the Bogoliubov
approximation. The
numerical calculations employ a computationally efficient procedure
based on a discrete variable representation (DVR) of the Hamiltonian. The DVR
is efficient for problems where the interaction potential may be expressed as a
local function of interparticle coordinates. In order to address condensates
that are both very large (millions of atoms) and fully anisotropic, the
ground state is found using a self-consistent field approach. Experience has
demonstrated, however, that standard iterative techniques applied to the
solution of the non-linear partial differential equation for the condensate are
non-convergent. This limitation is overcome using the method of direct
inversion in the iterated subspace (DIIS). In addition, the sparse structure
of the DVR enables the efficient application of iterative techniques such as
the Davidson and/or Lanczos methods, to extract the eigenvalues of physical 
interest. The results are compared with recent experimental data obtained 
for Bose-Einstein condensed alkali metal vapors confined in magnetic traps.
\end{abstract}
]

\narrowtext

\section{INTRODUCTION}

The experimental achievement of Bose-Einstein condensation (BEC) in dilute
alkali-metal gases confined in magnetic traps\cite{Cornell,Ketterle,Hulet,Hau}
has generated tremendous interest in the behavior of the inhomogeneous,
weakly interacting, and dilute Bose gas. At low temperatures, the confined
Bose gases have been shown to be well-described by mean-field linear-response
theories (MFLRT) based on the Bogoliubov\cite{Bogoliubov,Fetter1}
approximation, which assumes that the number of condensate atoms $N_0$ is a
substantial fraction of the total
number~\cite{BEC,Stringari0,Stringari1,Wallis,Smerzi,Sinha,Shi1}. The
finite-temperature
extensions of this theory, such as the Hartree-Fock-Bogoliubov and Popov
approximations\cite{Popov,Griffin1}, also have been successfully applied to
these systems\cite{Stringari2,Griffin2,Griffin3,Davies,You1,Shi2,Dodd}, though
the microscopic basis for these approaches and their agreement with experiment
remain somewhat uncertain\cite{Dodd,Burnett,Stoof}.

The usual approach taken within MFLRT is to first solve the (time-independent)
non-linear Schr\"odinger equation for a given number of atoms in the
condensate; the resulting wavefunction and chemical potential are then used in
the linear(ized) equations for the quasiparticle excitations in order to
obtain the eigenmodes of the system. At finite temperatures, the significant
depletion of the condensate requires that this procedure be iterated to
self-consistency. The magnitude of the non-linear, or self-energy, term
appearing in the Schr\"odinger equation for the ground state is proportional
to the condensate density. Thus, a fully three-dimensional solution of the
non-separable equation for the condensate presents a number of numerical
challenges, particularly for large values of $N_0$.

The magnetic trap used to confine these gases is well approximated by a
harmonic potential, in which the frequencies
($\omega_{x},\omega_{y},\omega_{z}$) are generally incommensurate. All of
the published experimental realizations to date have employed traps with either
spherical or cylindrical symmetry, which are more conducive to numerical 
study. The number of experimental groups which have obtained BEC of confined
alkali-metal gases is steadily rising, however\cite{homepage}. In order to
consider a range of possible trap geometries, as well as to permit future
investigations of any time-dependent properties, it is necessary to generalize
the numerical calculations to allow for complete anisotropy. The purpose of the
present paper is to demonstrate that the numerical obstacles may be overcome
and to outline a robust and computationally efficient procedure within the
MFLRT.

The approach taken by us is based on a discrete variable approximation
(DVR)\cite{Light} to the equations governing the statics of the condensate.
The DVR has several advantages over the methods used by others for the current
problem. Most standard techniques discretize the solution to the partial
differential equations (PDE) for the condensate and excitations either in
physical (grid) space or in function (basis set) space.  Grid-based
methods~\cite{grid} have the advantage that all local interactions between
particles have a local representation while the kinetic energy takes on a
sparse and typically banded structure. This sparsity is a consequence of
(typically) low-order finite difference approximations to the derivatives in
the PDE, and is crucial for the implementation of iterative
techniques~\cite{latext,iterative} developed for large, sparse linear
systems\cite{sparse}. The disadvantage of grid methods is that it is often
difficult to approximate the derivatives that appear in the PDE's to
sufficient accuracy without resorting to high-order differences or very small
step sizes. While an accurate representation of the kinetic energy is arguably
of minor importance for the ground state when $N_0$ is large, it is crucial for
the subsequent determination of excitations. Expansions in function space
(often called spectral and/or pseudospectral methods~\cite{spectral}) are
typically superior since it is possible to ``analytically'' differentiate the
functions without further approximation. These methods, however, require the
evaluation of potential energy matrix elements by quadrature, leading to a
non-diagonal and dense matrix representation of local operators. 
\par
The DVR exploits the dual relationship between certain orthogonal polynomials 
(such as the classical orthogonal polynomials) and the points and weights of a
Gauss quadrature. By using an appropriate set of polynomials as a basis for
expanding the solution to the PDE, it is 
possible to maintain almost all of the advantages of a grid-based approach 
as well as the global convergence of a finite basis set. In the DVR, any local
potential energy operator is diagonal and therefore easy to compute. The 
multidimensional kinetic energy also has a sparse representation because it is
a commuting sum of one-dimensional operators. The sparseness is not as
structured as is the case for more traditional approaches, but the expense of
generating the kinetic energy matrix elements is mitigated by the iterative
method used to solve the PDE: these terms need only be evaluated once. It is
also worth noting that the expansion coefficients of the solutions to the PDE
are trivially related to the values of the solution on the quadrature points
which serve as the physical grid. Perhaps the most important point to stress,
though, is that the solutions of the PDE scale very efficiently with basis set
size. This will become very important for three-dimensional problems where the
size of the various matrices could easily be as large as 100\,000 by 100\,000.
\par
The usual starting point for the theoretical treatment of the 
inhomogeneous, weakly interacting, dilute Bose gas at zero temperature is the
time-independent MFLRT in the Bogoliubov approximation. Since the derivation
of the resulting equations may be found in many places\cite{Fetter2}, only the
central results are presented here. In the absence of inhomogeneities other
than the confining potential, all wavefunctions may be assumed to be real. In
the weakly interacting and low-density limit relevant to the experiments of
interest, the interatomic interactions may be approximated by a two-body
delta-function contact potential with a coupling constant $g$. Minimizing the
grand canonical potential for interacting bosons and then linearizing the
resultant equations for the amplitude of the condensate $\psi({\bf r})$ and 
the excitations, $u({\bf r})$ and $v({\bf r})$, leads respectively to the 
Gross-Pitaevskii (GP)\cite{Gross,Pitaevskii} for the condensate and the 
Bogoliubov\cite{Bogoliubov,Fetter1} equations for the excitations:
\begin{equation}
\hat{L}\psi({\bf r})=\mu\psi({\bf r});\label{gp}
\end{equation}
\begin{eqnarray}
\left(\hat{L}-\mu+V_{\rm H}\right)u_n({\bf r})-V_{\rm H}v_n({\bf r})
&=&\epsilon_nu_n({\bf r});\label{bog1} \\
\left(\hat{L}-\mu+V_{\rm H}\right)v_n({\bf r})-V_{\rm H}u_n({\bf r})
&=&-\epsilon_nv_n({\bf r}),\label{bog2}
\end{eqnarray}

\noindent where the non-linear Schr\"odinger operator is written as
\begin{equation}
\hat{L}=-{\hbar^2\over 2M}\vec{\nabla}^2+V_{\rm trap}({\bf r})+V_{\rm H}.
\end{equation}
The non-linearity of the GP equation is due to the mean-field, or Hartree, 
potential $V_{\rm H}=g|\psi({\bf r})|^2$; at long wavelengths, the coupling
constant $g\approx 4\pi\hbar^2a/M$, where $a$ and $M$ are respectively the
$s$-wave scattering length and the mass of the atom. Only the case $a>0$ will
be considered here. The chemical potential, $\mu$, fixes the number of atoms
$N_0$ in the condensate (the contribution from the excitations is ignored in
the Bogoliubov approximation), and $\epsilon_n$ are the collective excitations
of the system. Note that the condensate is the zero-energy solution of the
Bogoliubov equations, $\psi({\bf r})=u_0({\bf r})=v_0({\bf r})$; the excitation
frequencies are measured with respect to the ground state energy $\mu$.

\par
The confinement due to the magnetic trap is well-described by a completely
anisotropic harmonic oscillator potential:
\begin{equation}
V_{\rm trap}={M\over 2}\left(\omega_x^2x^2+\omega_y^2y^2+\omega_z^2z^2\right).
\end{equation}

\noindent The analysis of the GP and Bogoliubov equations is considerably
simplified by re-expressing the trap frequencies $\omega$ and coordinates in
terms of the scaled variables
$(\omega_x,\omega_y,\omega_z)=\omega_0(1,\alpha,\beta)$ and
$(x,y,z)\rightarrow d_0(x,y,z)$, where $d_0=\sqrt{\hbar/M\omega_0}$ is the
characteristic oscillator length in the $\hat{x}$-direction. As a result, the
coupling constant becomes $g\rightarrow 4\pi\eta_0$, where we define
$\eta_0\equiv N_0a/d_0$. The Schr\"odinger operator is then
\begin{equation}
\hat{L}=-{1\over 2}\vec{\nabla}^2+{1\over 2}\left(x^2+\alpha^2y^2+\beta^2z^2
\right)+4\pi\eta_0\psi^2({\bf r}).
\label{schrodinger}
\end{equation}

\noindent All energies (including $\mu,\epsilon_n$) are now given in trap units
$\hbar\omega_0$. With these choices, the condensate and excited-state
wavefunctions are normalized to unity in the rescaled (dimensionless)
coordinates:
\begin{eqnarray}
\int d{\bf r}\;\psi^2({\bf r})&=&1;\label{normal_psi} \\
\int d{\bf r}\;\left[u_n^2({\bf r})-v_n^2({\bf r})\right]&=&1.
\label{normal_bog}
\end{eqnarray}

\noindent In the next section we discuss the numerical methods used to solve
these equations.
\section{NUMERICAL METHODS}
\par
The discrete variable representation is particularly useful for the problem
of trapped Bose-condensed atoms. As discussed in the Introduction, all
potential energy operators in the Hamiltonian are local and therefore have a
diagonal matrix representation, while the kinetic energy still has a fairly
simple and sparse structure. Since only the low-lying modes are relevant to
the static properties of the weakly interacting Bose gas at low temperatures,
the sparse Hamiltonian matrices are ideally suited to iterative techniques for
the determination of eigenvalues. Since these methods are dominated by
matrix-vector multiplies, a structured and sparse matrix offers considerable
computational savings.
\par
An important feature of the GP equation is that the interaction potential
depends non-linearly on the solution. Thus, for a given $N_0$, the lowest
eigenvalue $\mu$ and eigenvector $\psi({\bf r})$ must be found
self-consistently. In practical terms, a robust iterative method is required.
Indeed, as discussed below, no solution to the GP equation may be found simply
by straight iteration when the universal scaling parameter
$\eta\equiv\alpha\beta\eta_0
\mathrel{\lower1pt\hbox{$>\atop\raise2pt\hbox{$\sim$}$}}10$. Since $a/d_0$
is typically of order $10^{-3}-10^{-2}$, only a few thousand atoms in the trap
(several orders of magnitude fewer than are experimentally relevant) would be
sufficient to prevent the solution to the GP equation by iterative methods. We
have developed a variant of a well-known technique, called direct
inversion in the iterative subspace (DIIS)\cite{Pulay} to render convergent
the self-consistency process for the large $N_0$ of practical interest. DIIS
complements alternative numerical approaches, such as the method of steepest
descents (imaginary-time propagation), which have been successfully applied to
the trapped Bose condensates~\cite{Dalfovo,Esry,Ohberg1}.
\par
\subsection{DVR Techniques}

The basic ideas of a discrete variable representation are quite old.  
The earliest applications~\cite{DickCert} were designed to simplify the
calculation of certain classes of matrix elements that appeared in finite
basis set, variational calculations.  Subsequent
authors\cite{Light,KosKos,MarsKurt,Wyatt,Mucker} began to use
the DVR more directly and, in some instances, to view it as the primary 
representation for the problem under consideration.  The viewpoint of the
current authors is that the DVR follows naturally from a particular choice of 
a finite basis set, one which is mathematically linked to Gauss quadratures.
\par
Let us consider a basis of functions, $\left\{ \phi_{n}(x), n=1,N \right\}$, 
perhaps satisfying some set of boundary conditions over a finite or infinite 
interval, which is complete enough for expanding any unknown function over 
that interval to sufficient accuracy.  For simplicity we assume that these 
functions form an orthonormal basis for the space.  What we seek is a 
complementary set of ``coordinate eigenfunctions'', 
$\left\{ u_{i}(x), i=1,N \right\}$, and a generalized quadrature consisting of 
roots and weights $\left\{ x_{k}; w_{k},k=1,N \right\}$, such that
\begin{equation}
u_{i}(x_{k}) = \delta_{i,k}.\label{delta}
\end{equation}
We now expand the unknown $u_{i}(x)$ in the set of functions $\phi_{n}(x)$,
\begin{equation}
u_{i}(x) = \sum_{n=1}^{N} \phi_{n}(x) \< \phi_{n} \mid u_{i} \>,
\end{equation} 
and assume it is possible to evaluate the overlap integral,
$\< \phi_{n} \mid u_{i} \>$, sufficiently accurately using the 
generalized quadrature rule so that
\begin{eqnarray}
 \< \phi_{n} \mid u_{i} \> = w_{i} \phi_{n}(x_{i});\label{overlap} \\
 u_{i}(x) = w_{i} \sum_{n=1}^{N} \phi_{n}(x) \phi_{n}(x_{i}).\label{x_funcs}
\end{eqnarray}
The result~(\ref{overlap}) follows directly from Eq.~(\ref{delta})
and the integration rule for a Gauss quadrature:
\begin{equation}
\<f\mid g\>\equiv\int_a^bdx\;w(x)f(x)g(x)\equiv\sum_kw_kf(x_k)g(x_k),
\label{rule}
\end{equation}

\noindent where $w(x)$ is a non-negative weight function. Moreover, the
delta-function property (\ref{delta}) of the coordinate eigenfunctions is
exactly satisfied by Eq.~(\ref{x_funcs}) at the quadrature points $x_k$, since
the $\phi_n$ form a complete set:
\begin{equation}
\sum_{n=1}^N\phi_n(x_k)\phi_n(x_i)={\delta_{i,k}\over w_i}.
\end{equation}

\noindent It should be underlined that the $u_i(x)$ defined by
Eq.~(\ref{x_funcs}) are highly localized in the vicinity of the quadrature
points but are not true delta-functions, since they are finite for $x\neq x_k$.
\par
Since the coordinate eigenfunctions $u_i(x)$ are defined as continuous 
functions of $x$, it is possible to differentiate and tabulate them using the
known properties of the $\phi_{n}(x)$. The crucial step, of course, is to be
able to find the set of basis functions and the related generalized
quadrature which justify Eqs.~(\ref{overlap},\ref{x_funcs}).
\par
While no conditions have thus far been placed on the set $\phi_{n}(x)$, other
than orthonormality and completeness, in practice the requirement that the
$u_{i}(x)$ satisfy Eq.~(\ref{delta}) and Eqs.~(\ref{overlap},\ref{x_funcs})
limits the basis functions to the classical orthogonal polynomials. These are
defined by a three term recursion relation of the form
\begin{equation}
\beta_{j}\phi_{j}(x) = ( x - \alpha_{j} )\phi_{j-1}(x) - 
                        \beta_{j-1}\phi_{j-2}(x),\label{recursion}
\end{equation}
with the properties
\begin{equation}
\<\phi_i\mid\phi_j\>=\delta_{i,j}\;,\qquad\phi_0(x)={\rm const.}
\end{equation}

\noindent where the inner product is defined in Eq.~(\ref{rule}). Since
Eq.~(\ref{recursion})
may be interpreted as the definition of the coordinate operator $x$ in the
polynomial basis, the orthonormal eigenvectors of this tridiagonal matrix
must diagonalize the coordinate operator\cite{Schneidpoly}. It is therefore
not a coincidence that the associated eigenvalues are the generalized Gauss
quadrature points associated with the original basis $\phi_{n}(x)$. For this
set of $N$ functions there is an associated generalized Gauss quadrature
consisting of $N$ points and weights which assure us that
Eqs.~(\ref{overlap},\ref{x_funcs}) are satisfied {\it exactly}. 
\par
There are as many ways to define the $u_{i}(x)$ as there are classical
orthogonal polynomials. The most natural choice, however, is to use the
Lagrange interpolating functions, which explicitly diagonalize
Eq.~(\ref{recursion}). The Lagrange polynomials may be defined at the 
associated Gauss quadrature points by:
\begin{equation}
v_i(x)={\prod_{k=1}^N}'{x-x_k\over x_i-x_k},\label{lagrange}
\end{equation}
\noindent where the prime denotes exclusion of the point $x_i$ in the product.
The Gauss-Legendre quadrature points $x_k$ are defined on the interval
[-1:1]; with the associated weights, one obtains the eigenfunctions:
\begin{equation}
u_{i}(x) = v_{i}(x)/\sqrt{w_{i}}. \nonumber
\end{equation}
These polynomials have the ``delta-function property'' by construction, and
are normalized so that they form an orthonormal set under the generalized
$N$-point Gaussian quadrature:
\begin{equation}
\<u_i\mid u_j\>=\delta_{i,j}.
\end{equation}
This definition ensures that there is an equality between exact integration
and integration by quadrature for any integrand which may be represented as a
polynomial of degree $(2N-1)$ or smaller.  In the present work, both Lagrange
functions as well as a DVR based on a Hermite polynomial basis are considered.

Since the $u_{i}(x)$ diagonalize the coordinate operator, the matrix element
of any operator ${\mathcal O}(x)$ which is a local function of $x$ satisfies
\begin{equation}
\<u_i\mid{\mathcal O}(x)\mid u_j\>=\delta_{i,j}{\mathcal O}(x_i).\label{diag}
\end{equation}

\noindent The DVR basis not only considerably simplifies the evaluation of
many matrix elements of the Hamiltonian, but leads to a sparse representation
as well.  For many large matrix problems the only practical methods of
diagonalization or matrix inversion require the operation of the Hamiltonian
matrix on some known vector.  Sparsity is a key ingredient to performing this
operation efficiently.

While the result (\ref{diag}) is an identity within a particular $N$th-order 
Gauss quadrature, it is only exact when the product of $u_{i}(x)$, $u_{j}(x)$ 
and the local operator ${\mathcal O}(x)$ is a polynomial of degree $2N-1$ or 
smaller~\cite{BayeHeenen}. The present formalism, therefore, is particularly
conducive to the solution of the GP equation in the limit of large $N_0$. In
this case, the contribution of the kinetic term to the total energy is small.
The square of the condensate wavefunction may then be written in the so-called
Thomas-Fermi (TF) approximation:
\begin{equation}
\psi^2({\bf r})\approx\left[\mu_{\rm\sc TF}-{\case1/2}(x^2+\alpha^2y^2
+\beta^2z^2)\right]/4\pi\eta_0,\label{tf}
\end{equation}

\noindent where the normalization condition (\ref{normal_psi}) yields the 
chemical potential in the TF limit:
\begin{equation}
\mu_{\rm\sc TF}={\case1/2}\left(15\alpha\beta\eta_0\right)^{2/5}
\equiv{\case1/2}\left(15\eta\right)^{2/5},
\label{mu_tf}
\end{equation}

\noindent in units of $\hbar\omega_0$.
In the TF limit, both the interaction and confining potentials
appearing in the Schr\"odinger operator $\hat{L}$ (\ref{schrodinger}) are 
evidently second-order polynomials. For a finite but large number of atoms,
however, there is a small exponential tail near the boundary of the condensate
cloud resulting from the finite kinetic energy\cite{Stringari3,Pethick,Feder}.
As shown in Sec.~\ref{sec:results}, this exponential behavior may be
effectively captured even when using a comparatively low-order basis.

For the present three-dimensional calculation, the condensate and excited-state
wavefunctions are expanded using Cartesian coordinates in a product basis of
coordinate functions for each dimension; in the condensate case, for example,
one writes:
\begin{equation}
\psi({\bf r })=\sum_{i,j,k}c_{ijk}u_{i}(x)u_{j}(y)u_{k}(z).
\label{expand}
\end{equation}

\noindent The components of the three-dimensional kinetic energy
$T_{x,y,z}\equiv -{\case1/2}\nabla_{x,y,z}^2$ has the following representation
in the DVR product basis:
\begin{eqnarray}
& &\< u_{i}u_{j}u_{k}\mid T\mid u_{l}u_{m}u_{n}\>\nonumber \\
& &\qquad =T_{i,l}\delta_{j,m}\delta_{k,n}+T_{j,m}\delta_{i,l}\delta_{k,n}
+T_{k,n}\delta_{i,l}\delta_{j,m}.
\end{eqnarray}

\noindent Thus, the Hamiltonian matrix separates into a sum of dense,
one-dimensional kinetic energy matrices times Kronecker delta functions in the
remaining variables, plus a purely diagonal matrix associated with the
potential terms. While multidimensional expansions of this kind (\ref{expand})
exist in other coordinate systems\cite{Hoffman,CoLe,CoTrLe,Lemoine},
the choice of a product basis in Cartesian coordinates yields a separable
kinetic energy operator which ensures a sparse matrix representation of the
Hamiltonian in the multidimensional DVR space. The expansion coefficients
$c_{ijk}$ found by diagonalization are proportional to the values of the
wavefunction at the appropriate quadrature points.

\subsection{Iteration and the DIIS Method}

The central numerical difficulty in the solution of the GP equation is the
non-linearity associated with the two-body interactions. Since the magnitude
of the Hartree potential is proportional to the number of atoms in the
condensate, the self-consistent solution when $N_0$ is large can become
problematic. Many calculations reported to date have ``inverted'' the search;
that is, the value of $\mu$ is fixed while the wavefunction and associated
value of $N_0$ are obtained which satisfy the appropriate boundary and
normalization conditions. Such a procedure is straightforward in one dimension
(though somewhat less so in two)\cite{BEC}, since a direct numerical
integration of the GP equation can be implemented. In three dimensions, a
root search procedure may be devised to accomplish the same thing, but the
calculation would be extremely time-consuming. With the method of steepest
descents, the solution of the three-dimensional GP equation could be obtained
by imaginary-time propagation. Though this approach has been shown to yield
accurate results for large numbers of atoms~\cite{Dalfovo,Esry,Ohberg1}, it
is not obvious that time-dependent techniques should be most suitable for the
solution of an eigenproblem.

An elegant and (as shown below) inexpensive method for the solution of the GP
equation is to fix $N_0$ and to self-consistently solve for $\mu$ and
$\psi({\bf r})$. Unfortunately, a direct iteration of Eq.~(\ref{gp}) does not
usually converge to the global energy minimum, even for relatively small
condensates ($N_0\sim 10^3$, $\eta\sim 1$). Initially, we attempted several
simple schemes in order to improve convergence. One approach was to use the TF
approximation (\ref{tf}) to begin the iteration sequence. This would seem to
be an excellent starting point when the condensate is large and the TF
solution is a reasonable approximation. In practice, we have found that the
poor behavior of the TF wavefunction near the condensate boundary did not, in
general, make this a viable procedure. Even correcting the TF result using a
boundary-layer perturbation approach\cite{Feder} did not seem to greatly
improve convergence.

Evidently, a robust numerical process is required which rapidly damps out the
errors as the iteration sequence proceeds. One rather crude approach is to use
a linear combination of the solutions from the $(i-1)^{\rm th}$ and
$i^{\rm th}$ steps to initiate the iteration process for the $(i+1)^{\rm th}$
step~\cite{You2,Pu}; the proper linear combination may be found by numerical
experimentation. The approach used by the present authors, known as Direct
Inversion in the Iterative Subspace (DIIS)\cite{Pulay}, is a more systematic
version of the above that uses the information from all of the previous
iterations. The DIIS method is well known in the quantum chemistry community,
where it is used to accelerate the convergence of self-consistent field
calculations. As discussed below, for small to intermediate numbers of atoms
($10^3\lesssim N_0\lesssim 10^6$), DIIS is found to be at least as
computationally efficient as the method of steepest descents. 

DIIS begins by defining an error which characterizes the convergence of
the iteration process. While the error $\hat{e}$ may be defined in any
number of ways, one that appears to be particularly stable numerically is
the commutator:
\begin{equation}
\hat{e}\equiv\left[\hat{H},\hat{\rho}\right],\label{commute}
\end{equation}

\noindent where $\hat{\rho}({\bf r},{\bf r}')$ is the density matrix 
$|\psi\>\<\psi|$.
Evidently, $\hat{e}$ vanishes at convergence. Note that both the density matrix
and the error are matrices having $S^{2}$ elements, where $S$ is the dimension
of the Hilbert space. The procedure is to expand the current expression for the
Hamiltonian and error as a linear combination of the $m$ previous values 
\begin{eqnarray}
    \hat{H}^{m+1}&=&\sum_{i=1}^ma_{i}\hat{H}^{i};\\
    \hat{e}^{m+1}&=&\sum_{i=1}^ma_{i}\hat{e}^{i},
\end{eqnarray}
subject to the constraint that
\begin{equation}
         \sum_{i=1}^ma_{i} = 1.\label{constraint}
\end{equation}
Minimizing the squared norm of $\hat{e}^{m+1}$ with the
constraint~(\ref{constraint}), one obtains $m+1$ linear equations:
\begin{equation}
     \sum_{j=1}^mB_{i,j}a_{j}-\lambda= 0,\label{linear}
\end{equation}
where
\begin{equation}
  B_{i,j}\equiv\<\hat{e}_i\mid\hat{e}_j\>,\label{bij}
\end{equation} 

\noindent and $j=1,2,\ldots,m$. The Lagrange multiplier $\lambda$ enforcing the
constraint yields the squared norm of the error.

The practical implementation of the DIIS algorithm, especially for large
Hamiltonian matrices, deserves some additional comment.  Although it appears
to be necessary to store the Hamiltonian matrix for each iteration, this is
not the case.  The only part of the Hamiltonian matrix that varies from 
iteration to iteration is the nonlinear potential, and this matrix is diagonal 
in the DVR. Inserting Eq.~(\ref{bij}) into (\ref{commute}), the matrix
elements $B_{i,j}$ are simply given by:
\begin{equation}
B_{i,j}=2\<\varphi_i\mid\varphi_j\>\<\psi_i\mid\psi_j\>
-2\<\varphi_i\mid\psi_j\>\<\psi_i\mid\varphi_j\>,\label{bij2}
\end{equation}

\noindent where $\varphi_i({\bf r})\equiv\hat{H}^i\psi_i({\bf r})$; it is
important to note that $\hat{H}^i$ is the Hamiltonian matrix constructed from
$\psi_{i-1}({\bf r})$, whose minimal eigenvector is $\psi_i({\bf r})$. On
each iteration, therefore, it is only necessary to store the solution vector as
well as the vector which represents the operation of the Hamiltonian matrix on
the solution vector. Since these objects are both of dimension $S$ rather than
$S^{2}$, they may easily be stored in central memory even on current
workstations as long as the number of DIIS iterations does not become too
large. 

To summarize, only the vectors $\psi({\bf r})$ and $\varphi({\bf r})$ need to
be stored at each iteration step. From these, the scalar products (\ref{bij2})
required to set up the set of linear equations (\ref{constraint}) and
(\ref{linear}) may be generated. Once the $m$ unknowns $a_i$ as well as
$\lambda$ have been found, a new nonlinear potential may be constructed from
the past guesses,
\begin{equation}
V_{\rm H}^{m+1}=\sum_{i=1}^ma_iV_{\rm H}^i,
\end{equation}

\noindent and the next self-consistent field cycle initiated. Furthermore, by
storing and not destroying all previous elements of $B_{i,j}$, only an
additional row and column need be computed at each step of the iteration.
Since the dimension of $B$ is assumed to be small, the additional storage
should not be a problem on most workstations.   

With the TF expression for the condensate wavefunction as the initial guess,
the DIIS algorithm yields a convergent solution for significantly larger
numbers of atoms ($N_0\sim 10^5$) than was possible with a direct
iterative procedure. As $N_0$ continues to grow, however, the number of DIIS
iterations to achieve convergence increases until eventually the process
fails. It appears that small errors in the intermediate solution of the GP 
equation, particularly in the `tail' region close to the condensate surface,
are amplified by the nonlinearity of the potential during the iteration
process. Combining DIIS with a more realistic initial guess, we have been able
to achieve convergence for the large number of atoms ($N_0\sim 10^6$) relevant
experimentally. Two schemes used to improve the starting approximation for the
Hartree potential have proved particularly valuable. By slowly increasing the
number of atoms, the converged solution for a slightly smaller condensate is
an excellent choice. Alternatively, one may solve a modified GP equation with
an exaggerated kinetic energy contribution, then slowly ramp down the degree of
exaggeration (this technique is similar in spirit to simulated annealing). A
simple modification of the original DIIS procedure was also
found to be quite helpful. The DIIS procedure is broken into cycles with some
maximum number of iterations allowed per cycle. At the end of each cycle, the
DIIS procedure is restarted using the best available solution. By numerical
experimentation it was found that once the root mean square error (or 
alternatively $\sqrt{\lambda}$) is reduced to $10^{-2}-10^{-3}$, it becomes
possible to restart DIIS and to rapidly reduce errors to $10^{-6}-10^{-7}$
with very few additional iterations. 

\subsection{Interpolation of the Wavefunction}
\label{interpolate}

As discussed above, in the limit of very large $N_0$ where the TF theory is
believed to be valid, both the interaction and confining potentials are
well-approximated by low-order polynomials in $x,y,z$. Since an $N$-point
Gauss quadrature is able to integrate a $(2N-1)$th-order function exactly, it
should be possible to capture the essential behavior of the condensate
wavefunction (squared) using only a very small number of quadrature points.
One is therefore left with a somewhat surprising conclusion: when the number
of atoms becomes very large, the resulting self-consistency problem should
simplify considerably. A very coarse DVR grid has two obvious advantages. The
reduced dimension of the Hamiltonian matrix would accelerate the solution of
the eigenproblem at each DIIS iteration. In addition, the fewer degrees of
freedom for spatial variations should enable the self-consistent solution for
the condensate to converge with fewer DIIS iterations. The crucial unknown is
whether the exponential tail at the condensate boundary, due to the finite
kinetic energy contribution, varies sufficiently rapidly to invalidate the
low-order approximation applicable in the TF limit.

Suppose that the gross features of the condensate wavefunction could be
found using a low-order Gauss quadrature, corresponding to only a few DVR
points in each spatial direction. The self-consistent solution obtained using
such a `coarse grid' would then make an excellent initial guess for a more
accurate `fine grid' calculation, if the interpolation between grids could be
implemented successfully. By increasing the number of DVR points in the mesh 
once or perhaps a few times, it should be possible to rapidly converge the
solution of the GP equation for virtually arbitrary numbers of atoms. The
interpolated wavefunction on the quadrature points $(x_l,y_m,z_n)$ associated
with the fine grid is approximately
\begin{equation}
\psi(x_l,y_m,z_n)\approx\sum_{ijk}c_{ijk}u_i(x_l)u_j(y_m)u_k(z_n),
\end{equation}

\noindent where the sum is taken over coarse-grid expansion coefficients and
coordinate eigenfunctions. Since the DVR points corresponding to the two
different Gauss quadratures are not generally coincident, the values of the
coarse-grid coordinate basis functions at the fine grid points must be
obtained explicitly using either Eq.~(\ref{x_funcs}) or (\ref{lagrange}).

\subsection{Extraction of Eigenvalues and Eigenvectors}

The direct extraction of all the eigenvalues and eigenvectors of
Eqs.~(\ref{gp}-\ref{bog2}) becomes computationally impractical for very large
matrices. In three-dimensional systems, a sizeable number of
basis functions (and therefore quadrature points) are usually required in order
to adequately represent the wavefunctions over all space. This is a
particularly important consideration for the excitations, whose rapid spatial
variations can only be captured by high-order polynomials. As a result, the
dimension of the matrices can become very large, even after
block-diagonalizing according to parity. A full diagonalization of the
eigenproblem, using standard and widely-available routines based on variants
of the Givens-Householder method, places considerable demands on both storage
and {\it cpu}-time. A full diagonalization is not necessary since only the
lowest eigenvalue ($\mu$) of the GP equation is required, and it is sufficient
to determine merely the lowest-lying excitations of the Bogoliubov equations
at zero temperature.

When the matrices are too large to fit in memory, iterative methods must be
used in order to extract the relevant low-lying states. These techniques
require the frequent operation of the Hamiltonian on a vector, and are
practical only if there are relatively few non-zero matrix elements. Indeed, a 
sparse representation of the Hamiltonian is a crucial feature of the DVR and
the separability of the kinetic energy operator. In
addition, standard iterative procedures return reasonable approximations only
to the extremal eigenvalues, in the spirit of a Rayleigh-Ritz variational
method. The GP equation~(\ref{gp}) yields a positive-definite, real, symmetric,
and sparse matrix whose lowest eigenvalue may be found using variants of the
well-known Lanczos \cite{Lanczos} or Davidson \cite{Davidson} algorithms.

The Davidson method, which is widely used by quantum chemists, develops a small
orthonormal subspace of vectors which is adequate to describe the eigenpairs
of physical interest.  The Hamiltonian is projected into this $p$-dimensional
subspace and the Rayleigh-Ritz variational principle ensures that an upper 
bound to the lowest $n$ eigenvalues is obtained from the process.  Since the 
computational procedure in the application of the Davidson algorithm may be 
unfamiliar to some readers, it is summarized below:

\begin{enumerate}
\item Choose an initial set of m orthonormal trial vectors ${\bf b}_{i}$\,,
where $m \geq n$.
\item Calculate ${\bf h}_{i}$ = $\hat{H}{\bf b}_{i}$, the effect of the 
      Hamiltonian matrix on these vectors.  
\item Calculate the Hamiltonian matrix $\<{\bf b}_{i}|{\bf h}_{j}\>$
      from the information generated in steps 1 and 2.
\item Solve the resultant small eigenvalue problem for the current values of
      the wavefunctions, ${\bf \Psi}_{q}^{\rm cur}$ and eigenvalues,
      $E_{q}^{\rm cur}$ by a direct method.
\item Calculate the residuals, 
\begin{equation}
  {\bf r}_{q} = \left(E_{q}^{\rm cur} - \hat{H} \right){ \bf \Psi}_{q}^{\rm cur}
\label{residual}
\end{equation} 
      and test for convergence.  If all roots are converged stop.  If not 
      select the unconverged residuals for further improvement.
\item Using the set of residuals solve the equation,
\begin{equation}
         \left(\tilde{\bf H } - E_{q}^{\rm cur} \right) {\bf \Phi_{q} } 
                           = {\bf r_{q}}\label{h_approx}
\end{equation}
      where $\tilde{\bf H}$ is some approximation to the exact $\hat{H}$.

\item Schmidt orthonormalize the ${\bf \Phi}_{q }$ to the ${\bf b}_{i}$ and
      then append them to the ${\bf b}_{i}$ to enlarge the vector space.  

\item Calculate ${\bf h}_{i}$ = $\hat{H}{\bf b}_{i}$, for the appended vectors
      and the additional matrix elements needed to border 
      $\< {\bf b}_{i}|{\bf h}_{j}\>$.
\item Return to step 4.
\end{enumerate}

In the original Davidson algorithm, which was designed for diagonally dominant
matrices, the approximate Hamiltonian used in step 6 was the diagonal. Then
the calculation of the solution to Eq.~(\ref{h_approx}) is trivial. For
diagonally dominant matrices, this Jacobi preconditioning method produces a
quickly converging set of new vectors, and the entire calculation is dominated
by step 2 in the sequence above.  Although the DVR has the desirable feature
that it produces a sparse representation, the matrix is unfortunately far from
being diagonally dominant. Without an alternative preconditioning technique,
the calculation can require several ($\sim S$) full cycles which becomes 
computationally prohibitive.
\par
Our approach to the preconditioning problem is based on using a separable
approximation to $\tilde{\bf H }$, i.e.\ a Hamiltonian which can be written as
a sum of commuting operators in the three-dimensional space. For simplicity,
we have chosen to use the bare trap Hamiltonian, but other more complicated 
choices can be made, subject to the separability requirement.
Eq.~(\ref{h_approx}) then becomes:
\begin{equation}
         \left( \tilde{\bf H}_{0}^{1} + \tilde{\bf H}_{0}^{2}
              + \tilde{\bf H}_{0}^{3} - E_{q}^{\rm cur} \right) {\bf \Phi}_{q} 
                           = {\bf r}_{q}.
\end{equation}
In order to solve this equation, we transform to the basis set which
diagonalizes the separable Hamiltonian, recognizing that the transformation
can be written as a product of three one-dimensional unitary transformations. 
Thus, the residual vector on the right hand side of Eq.~(\ref{residual}) is
transformed from the DVR to the diagonal representation, an operation which
scales as the size of the three-dimensional basis set times the one-dimensional
basis set.  This linear scaling with basis set size is a consequence of the
separable form of the approximate Hamiltonian, yet preserves the most
desirable features of the DVR.  Once the residual has been transformed to
the diagonal representation, Eq.~(\ref{h_approx}) is trivially solved and then
the solution vector is transformed back to the DVR representation. This
procedure is analogous to using fast Fourier transforms (FFT) to solve
multidimensional PDE's. The FFT is more efficient (but less general) due to
special character of the Fourier transformation but the separability is a key
factor to its properties.  By using this separable preconditioning, we have
been able to reduce the number of Davidson iterations to manageable size. In
addition, the vectors from a previous calculation for a smaller condensate
are found to be excellent trial vectors to initiate the calculation. The
result is a robust procedure which is quite efficient for large matrices. 
\par
The matrix associated with the Bogoliubov equations~(\ref{bog1},\ref{bog2}), 
in contrast, is non-symmetric and has real eigenvalues occurring in 
positive-negative pairs. This matrix must be addressed using routines such as
those based on an Arnoldi or a non-symmetric Davidson approach.  Since the 
low-lying excitations of the Bogoliubov equations are those closest
in magnitude to zero (relative to the chemical potential) and are therefore the
{\it least} extremal, it is most convenient to decouple the original equations
using the following linear combinations:
\begin{eqnarray}
f_n({\bf r})&\equiv&u_n({\bf r})+v_n({\bf r});\\
g_n({\bf r})&\equiv&u_n({\bf r})-v_n({\bf r}).
\end{eqnarray}

\noindent The resulting eigenproblem then becomes
\begin{equation}
\left(\hat{L}-\mu+2V_{\rm H}\right)\left(\hat{L}-\mu\right)f_n({\bf r})
=\epsilon_n^2f_n({\bf r}),\label{bog3}
\end{equation}

\noindent where $\epsilon_0=0$ corresponds to the ground-state energy relative
to the chemical potential. Once the functions $f_n$ are found, the left
eigenvectors $g_n$ may be obtained directly by utilizing
\begin{equation}
g_n({\bf r})=\epsilon_n^{-1}\left(\hat{L}-\mu\right)f_n({\bf r})
\end{equation}

\noindent subject to the normalization condition $\<f|g\>=1$, obtained from
Eq.~(\ref{normal_bog}). While the resulting eigenproblem remains non-symmetric
(since the condensate density does not commute with the operator $\hat{L}$),
the dimension of the matrix has been reduced by a factor of two. Furthermore,
the eigenvalues $\epsilon_n^2$ are positive-semidefinite and
real~\cite{Fetter1}. In practice, the composite operator on the left side of
Eq.~(\ref{bog3}) is never determined explicitly; doing so at each DIIS
iteration not only would be time-consuming, but also would lead to a dense
matrix. Rather, at each Arnoldi/Davidson iteration the vector is multiplied in
turn by the two distinct sparse operators.

\section{results and discussion}
\label{sec:results}

The techniques discussed in the previous section are particularly useful under
two circumstances: when the potential is completely anisotropic, and when the
number of condensate atoms is very large. For illustrative purposes, results
are presented for Bose-condensed sodium atoms confined in traps with three
different geometries, all of which have been realized experimentally. The
completely anisotropic trap considered here is a TOP trap with angular
frequencies in the natural ratio~\cite{jila_comment}
$(\omega_x,\omega_y,\omega_z)=\omega_0^{\rm A}(1,\sqrt{2},2)$ where
$\omega_0^{\rm A}=354\pi$~rad/s; BEC in such a system has been recently
observed by Phillips {\it et al.}\cite{Bill}. Other geometries considered are
the cigar-shaped Ioffe-Pritchard trap of W.~Ketterle\cite{Ketterle2} with
cylindrical symmetry and
$(\omega_x,\omega_y,\omega_z)\approx\omega_0^{\rm C}(1,13.585,13.585)$, where
$\omega_0^{\rm C}\approx 33.86\pi$~rad/s, and the approximately spherical
`4D' potential of L.~Hau\cite{Hau} with $\omega_0^{\rm S}\approx 87$~rad/s.
All the calculations assume an $s$-wave scattering length for sodium of
$a=(52\pm 5)a_0$, where $a_0$ is the Bohr radius\cite{Eite}.

\subsection{Condensate}

The solution of the GP equation~(\ref{gp}) on the coarse and fine grids
employed 60 and 180 harmonic oscillator basis functions (Hermite polynomials
with a Gaussian prefactor) in each spatial direction, respectively. Coordinates
were rescaled through $\{x,y,z\}\to\{x,y/\alpha,z/\beta\}$; the non-linear
coefficient becomes $g=4\pi\eta=4\pi\alpha\beta\eta_0$ in order to ensure the
proper normalization of
the condensate density~(\ref{normal_psi}). With this choice, the condensate
cloud becomes almost spherical at large particle numbers, with an approximate
radius given by the TF expression $R_{\rm TF}=(15\eta)^{1/5}$, in units
of the characteristic trap length $d_0=\sqrt{\hbar/M\omega_0}$. DVR points
significantly beyond the TF radius were ignored. The ground state was assumed
to have totally even parity, so the GP equation was solved in a single
octant. The basis functions were eigenfunctions of a trap with frequencies
reduced from their actual values by a factor of 20, in order to decrease the
density of points. The resulting Hilbert spaces for the coarse and fine grids
then have dimension $3\,566$ and $18\,685$ respectively, for all geometries
and condensate numbers; both sizes are considerably reduced from the impractical
values of $216\,000$ and $5\,832\,000$ one naively might have obtained.

In Table~\ref{muse}, the chemical potentials obtained numerically using the
coarse grid are given as a function of the number of atoms $N_0=2^q$ in the
condensate
for the spherical, cylindrical, and anisotropic geometries described above.
The TF values, from Eq.~(\ref{mu_tf}), are included for comparison. The	number
of DIIS iterations required to yield a convergent solution to the GP equation
increases with the number of atoms; in the cylindrical case, DIIS failed to
converge for $N_0=2^{19}=524\,288$ and greater. Choosing a finer grid always
reduces the number of DIIS iterations; although increasing the number of points
yields additional degrees of freedom for the Hartree potential, the variations
of the condensate density (particularly in the surface region) are better
captured. In the cylindrical case, which has a cigar shape, the coarse grid has
too few DVR points in the two strongly-confining directions to adequately
capture the behavior of the condensate at large $N_0$. For $q>18$, therefore,
initial grids were chosen to contain a larger number of points not exceeding
$10^4$. Since more points imply a larger eigenproblem at each iteration,
practical calculations require as coarse a grid as possible.

\begin{figure}
\centerline{\psfig{figure=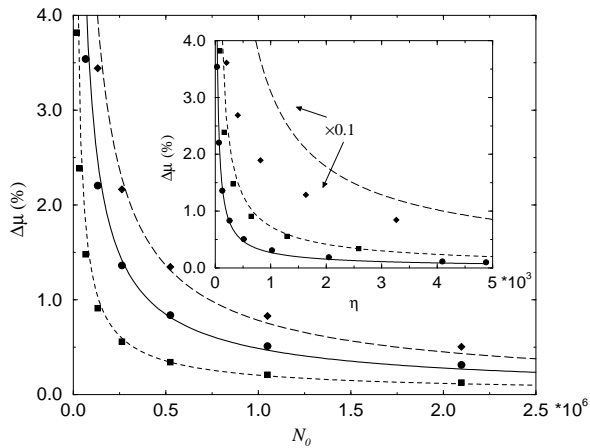,width=0.9\columnwidth,angle=270}}
\caption{The percent difference between the numerical and TF values for the
chemical potential $\Delta\mu=1-\mu_{\rm TF}/\mu_{\rm exact}$ are shown as a
function of the number of atoms and the universal scaling parameter
$\eta=\alpha\beta(N_0a/d_0)$ (which is proportional to the non-linear coupling
constant) for all three trap geometries. The data for the 
spherical (circles), cylindrical (diamonds), and anisotropic (squares) cases
are identical to those given in Table~\ref{muse}. The solid, dashed, and
dotted lines are fits to the data points using the expression
$\Delta\mu\approx\gamma^{\{\rm S,C,A\}}/\mu^{\{\rm S,C,A\}}$ for the spherical
($\gamma^S=1.5$), cylindrical ($\gamma^C=170$), and anisotropic
($\gamma^A=4.0$) cases, respectively. It is important to note that in the
inset, the values of $\Delta\mu$ for the cylindrical data are reduced by a
factor of 10 in order to facilitate comparison.}
\label{mu_fig}
\end{figure}

The convergence of the chemical potentials with the number of condensate atoms
is shown in Fig.~\ref{mu_fig}. Both the cylindrical and spherical traps give
rise to relatively slow convergence of the chemical potential to the TF value
compared with the anisotropic case. This trend is expected for the spherical
trap, where the confinement is extremely weak. In the cylindrical case the
condensate is strongly confined, but only in the radial direction where the
motion of the atoms is more or less frozen. Since the axial kinetic energy can
be large, however, the cylindrical trap is effectively loose and
one-dimensional. In contrast, the fully anisotropic trap considered here is
relatively tight in all directions, and the chemical potential converges to
the TF value more rapidly.

In the TF limit, most relevant quantities, such as the mean condensate radius
$R_{\rm TF}=(15\eta)^{1/5}d_0$ and the chemical potential
$\mu_{\rm TF}={\case 1/2}(15\eta)^{2/5}\hbar\omega_0$, are functions of the
universal scaling parameter $\eta=\alpha\beta(N_0a/d_0)$. Similarly, the
first-order correction to the TF chemical potential $\mu_{\rm TF}$, taking
into account the average kinetic energy~\cite{Stringari3,Pethick} and
contributions from the potential energies~\cite{Feder}, is proportional to
$R_{\rm TF}^{-2}\sim\mu_{\rm TF}^{-1}$. The fits to the numerical data of
$\Delta\mu\equiv(1-\mu_{\rm TF}/\mu)\approx\gamma/\mu_{\rm TF}$, shown in
Fig.~\ref{mu_fig}, are in reasonable agreement with this behavior (except for
the cylindrical case for very small numbers of atoms $N_0\lesssim 10^4$). One
might naively expect, therefore, that the convergence of the chemical
potential to the TF limit as a function of $\eta$ (which is proportional to
the coefficient of the non-linear term in the GP equation) would be
independent of trap
geometry. As shown in Fig.~\ref{mu_fig}, this is in fact not the case.
The data for the cylindrical case converge far more slowly than those for the
other geometries, though the magnitudes of the TF chemical potentials for a
given value of $\eta$ are identical (note that the values of $\Delta\mu$ shown
in the figure for the cylindrical case are reduced from their actual values by
an order of magnitude in order to facilitate comparison). The kinetic energy
contribution to the chemical potential (and therefore to the total energy) is
evidently strongly dependent on trap geometry. Thus, a large non-linear term
in the GP equation does not necessarily mean that the TF approximation 
adequately represents the system.

In order to verify the accuracy of the solution obtained on the coarse grid,
the condensate wavefunction was interpolated onto a $18\,685$-point grid
derived from a DVR basis with 180 polynomials in each direction (refer to 
Sec.~\ref{interpolate} for details on the interpolation technique). The
solution of the GP equation was again converged on the fine grid. For small to 
intermediate values of
the nonlinear coupling ($\eta\lesssim 3000$), a single interpolation between
the two meshes was sufficient to yield a solution on the fine grid with only a
few ($\sim 10$) additional iterations. As $\eta$ increased, two or more
successive interpolations and re-convergences were generally required before
the solution on the finest grid could be obtained. In all cases, the values of
the chemical potential at the two extremes were found to be identical to at
least three decimal places.

\begin{figure}
\centerline{\psfig{figure=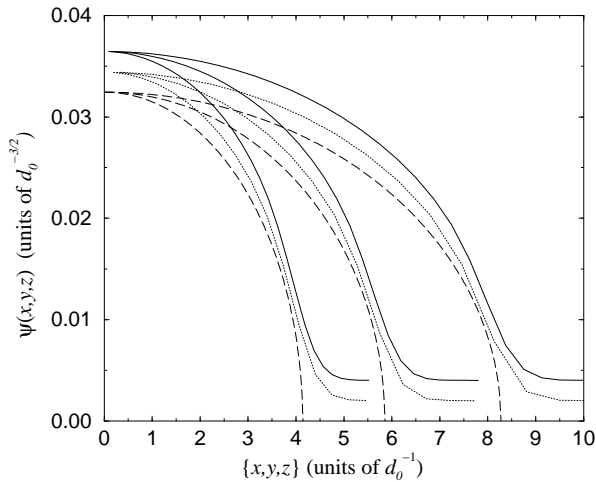,width=0.9\columnwidth,angle=270}}
\caption{The condensate wavefunction for $N_0=2^{19}=524\,288$ atoms in the
fully anisotropic trap, normalized to unity, is shown as simultaneous
projections along the positive $\hat{x}$ (rightmost curves), $\hat{y}$ (middle
curves), and $\hat{z}$ (leftmost curves) axes. The dashed lines correspond to
the TF approximation. The numerical results obtained on the coarse and fine
grids are shown as dotted (offset $0.002$) and solid (offset $0.004$) lines,
respectively. The interpolated coarse-grid and converged fine-grid
wavefunctions exactly coincide.}
\label{psi_fig}
\end{figure}

The condensate wavefunctions obtained on the coarse and fine grids are compared
in Fig.~\ref{psi_fig}, for the case of $N_0=2^{19}=524\,288$ atoms in the
completely anisotropic trap. The coarse-grid condensate profile along a given
axis appears rather crude, particularly in the surface region where the
wavefunction varies rapidly; for a given basis, convergence is only obtained
at the DVR points. Nevertheless, this wavefunction interpolated is
virtually indistinguishable from the converged solution on the fine mesh. As
$\eta$ increases, however, the length of the tail at the surface of the
condensate shortens. Eventually, the coarse grid will have too few points in
this crucial region to adequately capture the rapid variations. In this case, a
large jump in mesh size results in an interpolated wavefunction that more
poorly represents the self-consistent result.

\begin{table}
\begin{tabular}{ccccc}
$q$ & $\mu^{\rm S}$ & $\mu^{\rm C}$ & $\mu^{\rm A}$ \\ \hline
0 & 1.500 & 14.085 & 2.207 \\
10 & 1.825 (1.119) & 17.384 (9.393) & 3.572 (2.824)\\
11 & 2.065 (1.477) & 19.392 (12.395) & 4.345 (3.726)\\
12 & 2.435 (1.949) & 22.359 (16.355) & 5.425 (4.917)\\
13 & 2.970 (2.571) & 26.620 (21.580) & 6.904 (6.488)\\
14 & 3.719 (3.393) & 32.682 (28.475) & 8.900 (8.560)\\
15 & 4.743 (4.477) & 41.055 (37.573) & 11.572 (11.296)\\
16 & 6.124 (5.907) & 52.433 (49.578) & 15.128 (14.904)\\
17 & 7.970 (7.795) & 67.750 (65.418) & 19.847 (19.667)\\
18 & 10.427 (10.285) & 88.228 (86.320) & 26.096 (25.950)\\
19 & 13.685 (13.571) & 115.453 (113.900) & 34.358 (34.241)\\
20 & 17.999 (17.907) & 151.545 (150.29) & 45.275 (45.182)\\
21 & 23.702 (23.628) & 199.313 (198.31) & 59.693 (59.618)\\
\end{tabular}
\caption{The chemical potentials $\mu^{\{S,C,A\}}$ corresponding to spherical,
cylindrical, and anisotropic geometries respectively, are given in units of
$\hbar\omega_0^{\{S,C,A\}}$ for various numbers of condensate atoms
$N_0\equiv 2^q$. The TF values, from Eq.~(\ref{mu_tf}), are given in
parentheses. All results are converged to three decimal places and were
obtained using the coarse grid with at least $60^3$ basis functions and
$3\,566$ DVR points.}
\label{muse}
\end{table}

\subsection{Excitations}

The excitations of a condensate in a fully anisotropic harmonic trap have been
completely classified~\cite{Fliesser}, and have been explicitly obtained in the
low-density~\cite{Marinescu} and TF~\cite{Ohberg2,Csordas} limits. The states
are polynomials of order $N=l+m+n$ and are labeled by the total parity
$P=(-1)^{l+m+n}$, where the quantum numbers $(l,m,n)$ represent the order of
the polynomials along the $(x,y,z)$-directions in the non-interacting limit.
In the strongly-interacting (or hydrodynamic) regime, there are four odd and
four even parity low-lying modes with energies
$\epsilon=\sqrt{l+\alpha^2m+\beta^2n}$ in
units of $\hbar\omega_0$, where $(l,m,n)$ can be either $0$ or $1$. The ground
state, relative to the chemical potential, has quantum numbers $(0,0,0)$. The
only states with $N=1$ are the odd-parity dipole modes, where the center of
mass oscillates with the three trap frequencies. For $N=2$, there are six
quadrupole oscillations with even parity and stationary center of mass. Three
of these have energies given by the expression above, with $(l,m,n)=(1,1,0)$
and cyclic permutations. The other three are the solutions of the secular
equation:
\begin{equation}
\left|\matrix{3-\epsilon^2 & 1 & 1\cr 1 & 3-\epsilon^2/\alpha^2 & 1\cr
1 & 1 & 3-\epsilon^2/\beta^2\cr}\right|=0.
\end{equation}

\noindent For $(\alpha,\beta)=(\sqrt{2},2)$, the geometry considered here, one
readily obtains $\epsilon=\sqrt{8\pm 4\sqrt{2}}$ and $\sqrt{5}$. In the fully
anisotropic case, all the excitation energies (except those for the odd-parity
dipole modes) decrease with $N_0$~\cite{Marinescu}.

\begin{figure}
\centerline{\psfig{figure=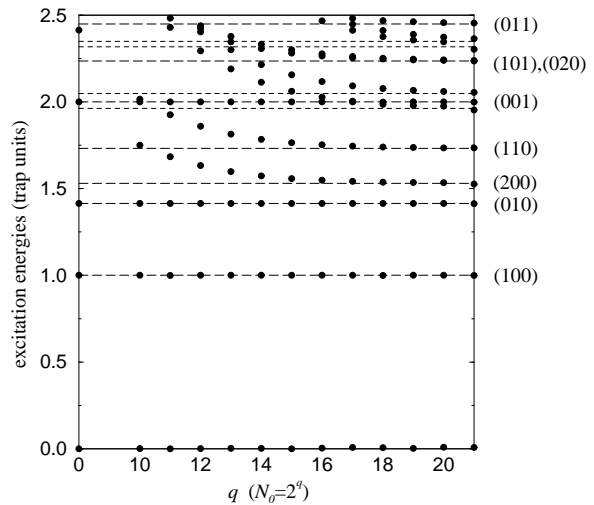,width=0.9\columnwidth,angle=270}}
\caption{The low-lying excitations of a condensate in the completely
anisotropic trap are given in trap units $\hbar\omega_0$ as a function of $q$,
where the number of atoms $N_0=2^q$. The circles correspond to numerical
results; horizontal dashed lines are the predictions of the TF theory. The
data points along zero are the ground-state energies relative to the chemical
potentials listed in Table~\ref{muse}. The long-dashed modes are labeled by
$(lmn)$, where $l$, $m$, and $n$ are the quantum numbers of the
non-interacting harmonic oscillator states along $x$, $y$, and $z$,
respectively. The unlabeled short-dashed excitations with energies just
below and above $2\hbar\omega_0$ are higher-order modes with odd parity along
$x$ and $y$, respectively; the two at higher energies have totally even parity
(lower) and odd parity in both $x$ and $y$ (upper).}
\label{exc_bill}
\end{figure}

The low-lying excitation energies for a completely anisotropic trap have been
computed numerically, and are shown in Fig.~\ref{exc_bill}. All the
calculations were obtained using the $3\,566$-point grid. While small
finite-size and coarse-graining effects are present, as evidenced by the small
fluctuations in the ground-state energy $\epsilon=0$, the results closely
match the TF predictions described above. As expected, the odd-parity dipole
modes are independent of the number of atoms in the trap. All the other
frequencies depend strongly on $N_0$, decreasing from their non-interacting
values in agreement with perturbative calculations~\cite{Marinescu}.

\begin{figure}[t]
\centerline{\psfig{figure=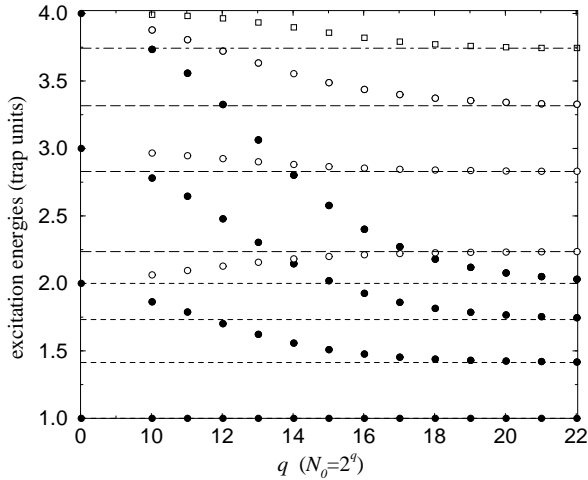,width=0.9\columnwidth,angle=270}}
\caption{Selected low-lying excitations $\epsilon_{nl}$ of the spherical
condensate are given in trap units $\hbar\omega_0$ as a function of $q$,
where the number of atoms $N_0=2^q$. The filled and open circles correspond
to numerical results for $n=0$ and $n=1$, respectively, while squares
represent $\epsilon_{20}$; horizontal dashed ($n=0$), long-dashed ($n=1$),
and dot-dashed ($n=2$) lines are the results obtained in the hydrodynamic
limit $\epsilon_{nl}=\sqrt{l+2n(n+l+3/2)}$.}
\label{exc_lene}
\end{figure}

While the frequencies of all the lowest-lying modes attain their large-number
values by $N_0\sim 10^6$ in the relatively strong anisotropic trap, the
convergence to the hydrodynamic limit slows as the quantum numbers increase and
the confinement is weakened. In Fig.~\ref{exc_lene}, the low-lying excitations
$\epsilon_{nl}$ of the loose spherical trap are shown as a function of the
number of atoms in the condensate, up to $N_0=2^{22}\approx 4.2\times 10^6$. In
the hydrodynamic limit, the energies are given by
$\epsilon_{nl}=\sqrt{l+2n(n+l+3/2)}$~\cite{Stringari0}.
The lowest number-dependent mode $\epsilon_{02}$ agrees with its hydrodynamic
value $\sqrt{2}$ to less than a percent by $N_0\approx 10^6$. The numerical
value of $\epsilon_{04}$, in contrast, differs from its limiting value of $2$
by approximately $2\%$ even when the number of atoms is as large as
$4\times 10^6$ atoms. Evidently, the magnitude of the excitation energy,
relative to the ground state $\mu$, does not alone provide a sufficient
indication of its convergence to the hydrodynamic limit. A similar
number-dependence for the higher-lying excitations of a
cylindrically-symmetric condensate has also been recently
obtained~\cite{You2}.

\section{conclusions}

A numerical procedure is introduced for the investigation of an
interacting Bose gas at zero temperature confined in a completely anisotropic
trap. The central feature of the technique is the use of the discrete variable
representation (DVR) as the primary basis for the calculations. The DVR
combines the best features of grid and basis-set techniques. All local
operators are diagonal, so the evaluation of interaction matrix elements
becomes trivial. While the kinetic energy has a more dense representation,
it may be evaluated analytically to high accuracy by exploiting the underlying
polynomial basis used to define the DVR. Furthermore, the kinetic energy needs
to be evaluated only once.

In the present method the condensate density is determined self-consistently;
for fully three-dimensional systems, this approach is considerably more
efficient than conventional root-search algorithms. At each iteration, the
ground-state wavefunction is obtained by iteratively diagonalizing the sparse
GP Hamiltonian, using either a Lanczos or Davidson method. Convergence of the
self-consistent solution to the GP equation is substantially hastened by
employing direct inversion in the iterated subspace (DIIS). As the non-linear
coefficient of the GP equation becomes very large, it often becomes necessary
to employ more sophisticated techniques, including number or kinetic energy
ramping and multigrid interpolation. In general, DIIS becomes more expensive
as the number of atoms increases; it is conceivable that an alternative method
such as imaginary-time propagation becomes more efficient than DIIS in the
regime $N_0\mathrel{\lower1pt\hbox{$>\atop\raise2pt\hbox{$\sim$}$}}10^6$.

The convergence of the chemical potential and the low-lying collective
excitations is investigated as a function of trap geometry. The chemical
potential is found to approach its Thomas-Fermi value more slowly as the 
confinement weakens or the degree of anisotropy becomes more pronounced; the
convergence does not scale universally with the magnitude of the non-linear
coefficient. The excitations of a completely anisotropic condensate have been
calculated numerically, and for large numbers of atoms agree with the
Thomas-Fermi predictions~\cite{Ohberg2,Csordas}. For a very weak spherical
trap, the collective frequencies converge to their hydrodynamic
values~\cite{Stringari0} more slowly.

\begin{acknowledgments}
The authors would like to acknowledge useful discussions with T. Bergeman,
C. W. Clark, R. J. Dodd, M. Edwards, A. L. Fetter, E. Hagley, W. D. Phillips,
and E. Tiesinga. This work was supported by the U.S. office of Naval Research.
\end{acknowledgments}


\begin{references}
\bibitem{jila_comment}
There are two natural configurations for the coils in a TOP trap, resulting
in oscillator spring constants either in the ratio $1:1:8$ (as in the JILA
experiments) or $1:2:4$.
\end{references}
\end{document}